\documentclass[useAMS,usenatbib]{mn2e}
\usepackage{amsmath}
\usepackage{graphicx}
\usepackage{txfonts}
\usepackage{natbib}
\usepackage{bm}

\def\beq#1{\begin{equation}\label{#1}}
\def\eeq{\end{equation}}
\def\beqa#1{\begin{eqnarray}\label{#1}}
\def\eeqa{\end{eqnarray}}

\def\Eq#1{Eq.~(\ref{#1})} 
\def\eqn#1{~(\ref{#1})}
\def\myfrac#1#2{\left(\frac{#1}{#2}\right)}

\def\comment#1{\relax}
\def\dfrac#1#2{\displaystyle\frac{#1}{#2}}
\def\mydfrac#1#2{\left(\displaystyle\frac{#1}{#2}\right)}

\title[Viscous-convective instability in Keplerian discs II]{A viscous-convective instability in laminar Keplerian thin discs. II. Anelastic approximation.}
\author[N. Shakura \& K. Postnov] {N. Shakura$^1$
\thanks{E-mail: nikolai.shakura@gmail.com, kpostnov@gmail.com},
K. Postnov$^{2, 1}$\\
$^1$ Sternberg Astronomical Institute, Moscow M. V. Lomonosov State University, Universitetskij pr., 13,  Moscow 119992, Russia\\
$^{2}$ Faculty of Physics,
M. V. Lomonosov Moscow State University,
Leninskie Gory, Moscow 119991, Russia\\}	

\begin{document}

\date{Received ... Accepted ...}
\pagerange{\pageref{firstpage}--\pageref{lastpage}} \pubyear{2012}

\maketitle

\label{firstpage}

\begin{abstract}
Using the anelastic approximation of linearised hydrodynamic equations, we
investigate the development of axially symmetric small perturbations in thin Keplerian 
discs. The sixth-order dispersion equation is derived and numerically solved for different values of relevant physical parameters (viscosity, heat conductivity, 
disc semi-thickness and vertical structure). The analysis reveals the appearance of 
two overstable modes which split out from the classical Rayleigh inertial modes 
in a wide range of the parameters in both ionized and neutral gases. 
These modes have a viscous-convective nature and can serve as a seed for 
turbulence in astrophysical discs even in the absence of magnetic fields.  

\end{abstract}

\onecolumn

\section{Introduction}
\label{intro}

In an attempt to understand hydrodynamic instabilities   which can 
potentially initiate turbulence in accretion discs, in \cite{2015MNRAS.448.3707S} (Paper I) we have performed a local WKB-analysis of axially symmetric perturbations
in thin accretion discs. It was found that under a special choice of wave
vector direction of the perturbations, almost 
(but not completely) aligned with the disc symmetry plane, the presence of a 
microphysical viscosity, parametrized in terms of the mean-free path
length to the disc scale ratio and taken into account in the dissipation function 
in the right-hand side of the energy equation, leads to the appearance of an overstable 
oscillating behaviour of one of two classical Rayleigh inertial modes. The 
instability was found in a wide range of perturbation wavelengths (expressed
through the dimensional wavelength vector $kr$, where $r$ is the disc radial scale),
around $kr\sim 30-100$, in both fully ionized gases and neutral gases. The microphysical heat conductivity was taken into account through the
dimensionless Prandtl number, Pr, which ranges from 0.052 for fully ionized plasma to 2/3 for neutral gases. 
We have found that the instability increment, reaching $\sim 0.1$
local Keplerian values, diminishes with decreasing the Prandtl number (e.g. due to 
the presence of a photon heat conductivity) and with increasing background
vertical entropy gradient (expressed in terms of the Brunt-V\"ais\"al\"a 
frequency).   
Such a behaviour of the instability is in agreement with physically intuitive dumping 
effect of the heat conductivity and entropy gradients on the development of small radial perturbations propagating under a small angle to the disc plane.    

To make the physics as simple as possible, in Paper I we have used the Boussinesq approximation of the hydrodynamic equations, which assumes the incompressibility of the fluid in the continuity equation and neglects the Euler pressure variations in the energy equation. We argued that the incompressibility approximation is justified 
for radial perturbations with $kr\gg 1$, which may suggest that the discovered viscous instability of the inertial Rayleigh modes is real and not the result of the approximations used. 
     
In this paper we continue studying the viscous-convective 
instability in thin shear laminar flows found in our paper \cite{2015MNRAS.448.3707S}. 
Here we treat the problem in the anelastic approximation and take into account 
the vertical boundary conditions in the thin Keplerian discs.
The anelastic  approximation is the next approximation to 
the full system of hydrodynamic equations, but in which the term $\partial \rho/\partial t=0$, i.e. the continuity equation 
takes the form 
$
div(\rho \bm{u})=0
$, which allows one to filter out sound waves. 

When considering sound-proof stratified flows, the use of the anelastic approximation 
is known to have some subtleties (see, for example, 
the recent analysis by \cite{2013ApJ...773..169V} and references therein). 
Special attention should be given to the energy equation, since the standard anelastic set of equations operates with adiabatic perturbations \cite{1962JAtS...19..173O}. 
Our analysis, in contrast, is heavily based on the viscous energy generation 
in the sheared flows, therefore the rigorous proof of the applicability 
of the anelastic approximation in this case is to be found. 
As a justification of this treatment we heuristically use 
the criterion that the linearised equations should not give rise to spurious   
modes with unphysical behaviour (e.g., unstable modes in the steady-state
solid-body rotation case).

Although our analysis is applicable for any sheared axially symmetric flow, 
we will be mostly concerned with thin Keplerian discs which have a wide range of 
phenomenological applications. This means that in the continuity equation of importance becomes the term $\sim 1/\rho_0(\partial\rho_0/\partial z)$, which can be quite significant in thin discs and which we have neglected in the Boussinesq approximation. The vertical boundary conditions in thin discs 
are taken into account by solving the Sturm-Liouville problem for the $z$-part
of perturbations. 

The main result of the paper is the dispersion equation 
\Eq{dispeq}, which is a sixth-order algebraic equation for small perturbations in the form $f(z)\exp(\mathrm{i}\omega t-k_rr))$. The solution of this equation 
signals the appearance of an overstable behaviour for two modes with the same 
negative imaginary part and real parts with equal absolute values but different signs which 
split from two classical inertial Rayleigh modes, in a wide range of $k_rr\gg 1$.

The structure of the paper is as follows. In Section \ref{s:basic}
we write down the basic equations. In Section \ref{s:linear} we linearise 
the full system of equation in the anelastic approximation.
We proceed with the derivation of the dispersion equation in Section \ref{s:dispeq}, 
following by its numerical analysis in Section \ref{s:solution}. Section \ref{s:discussion} summarizes our findings.   
 
\section{Basic equations}
\label{s:basic}

The system of hydrodynamic equations reads:

\begin{enumerate}
\item{mass conservation equation}
\beq{e:mass}
\frac{\partial \rho}{\partial t}+\nabla\cdot(\rho\bm{u})=0\,,
\eeq
In cylindrical coordinates for axially symmetric flows:
\beq{e:div}
\nabla\cdot(\rho\bm{u})=\frac{1}{r}\frac{\partial (\rho r u_r) }{\partial r}
+\frac{\partial (\rho u_z) }{\partial z}
\eeq

\item{Navier-Stokes equation including gravity force} 
\beq{e:NS}
\frac{\partial \bm{u}}{\partial t}+(\bm{u}\nabla)\cdot\bm{u}=-\frac{1}{\rho}\nabla p -
\nabla \phi_g +\bm{{\cal N}}\,.
\eeq
Here $\phi_g=-GM/r$ is the Newtonian gravitational potential of the central body with mass $M$, $\bm{{\cal N}}$ is the viscous force. In cylindrical coordinates for axially symmetric flows:
\beq{e:NSr}
\frac{\partial u_r}{\partial t}+ u_r \frac{\partial u_r}{\partial r}+
u_z\frac{\partial u_r}{\partial z}-\frac{u_\phi^2}{r}=
-\frac{\partial \phi_g}{\partial r}-\frac{1}{\rho}\frac{\partial p}{\partial r}+{\cal N}_r\,,
\eeq

\beq{e:NSphi}
\frac{\partial u_\phi}{\partial t}+ u_r \frac{\partial u_\phi}{\partial r}+
u_z\frac{\partial u_\phi}{\partial z}+\frac{u_r u_\phi}{r}={\cal N}_\phi\,,
\eeq

\beq{e:NSz}
\frac{\partial u_z}{\partial t}+ u_r \frac{\partial u_z}{\partial r}+
u_z\frac{\partial u_z}{\partial z}=
-\frac{\partial \phi_g}{\partial z}-\frac{1}{\rho}\frac{\partial p}{\partial z}+{\cal N}_z\,.
\eeq
The linearised viscous force components are specified in Appendix A of Paper I. 

\item{energy equation} 
\beq{e:en}
\frac{\rho {\cal R} T}{\mu}\left[
\frac{\partial s}{\partial t}+(\bm{u}\nabla)\cdot s
\right]=
Q_\mathrm{visc}-\nabla\cdot\bm{F}\,.
\eeq
where $s$ is the specific entropy per particle, 
$Q_\mathrm{visc}$ is the viscous dissipation rate per unit volume, 
${\cal R}$ is the universal gas constant, 
$\mu$ is the molecular weight, $T$ is the temperature, and terms on the right 
stand for the viscous energy production and the heat conductivity energy flux $\bm{F}$, 
respectively. The energy flux due to the heat conductivity is 
\beq{e:kappat}
\nabla\cdot\bm{F}=\nabla(-\kappa\nabla T)=-\kappa\Delta T-\nabla\kappa\cdot\nabla T\,.
\eeq
Note that both electrons and photons, and
at low temperatures neutral atoms, can contribute to the heat conductivity
(see Section \ref{s:solution} below).

\item{equation of state}

The equation of state for a perfect gas is convenient to write in the form:
\beq{e:eos}
p=Ke^{s/c_V}\rho^\gamma\,,
\eeq 
where $K$ is a constant, $c_V=1/(\gamma-1)$ is the specific volume heat capacity 
and $\gamma=c_p/c_V$ is the adiabatic index (5/3 for the monoatomic gas). 
We will also use the equation of state in the form
\beq{e:eos1}
p=\frac{\rho{\cal R}T}{\mu}\,,
\eeq
where $\mu$ is the molecular weight. 
\end{enumerate}

\section{linearised equations in anelastic approximation}
\label{s:linear}

The perturbed hydrodynamic variables can be written in the form $x=x_0+x_1$, 
where $x_0$ stand for the unperturbed quantities and $x_1=(\rho_1,p_1,u_{r,1},u_{z,1},u_{\phi,1})$ are small perturbations.
In contrast to Paper I in which we considered the local WKB approximation, i.e. small perturbations of density, pressure and velocity in the form 
$x_1(t,z,r)\propto \exp(\mathrm{i}\omega t-\mathrm{i}k_rr-\mathrm{i}k_zz)$, 
here we will take them in the form $\propto f(z)\exp(\mathrm{i}\omega t-\mathrm{i}k_rr)$ 
with the boundary conditions $f(z_0)=0$, $f(-z_0)=0$, where $z_0$ is the disc 
semi-thickness. We will consider thin discs with $z_0/r\sim u_s/u_{\phi,0}\ll 1$
($u_s$ is the sound velocity). Below we shall omit subscript $1$ for small
perturbations of the velocity, unless stated otherwise.


In this Section we will formultae the so-called anelastic approximation 
of hydrodynamic equations in which   
the sound wave perturbations are neglected 
by omitting the term $\partial \rho/\partial t$ in the 
continuity equation \cite{1962JAtS...19..173O}.

The linearised hydrodynamic equations are written as follows.
\begin{enumerate}
\item{} \textbf{Continuity equation}
  
The anelastic approximation for gas velocity $\bm{u}$ is $\nabla\cdot \rho_0 \bm{u}=0$:
\beq{e:anel}
\frac{\partial u_z}{\partial z}-\mathrm{i}k_r u_r+\frac{1}{\rho_0}\frac{\partial \rho_0}{\partial z}u_z
+\frac{1}{\rho_0}\frac{\partial \rho_0}{\partial r}u_r=0\,.
\eeq

\item{} \textbf{Dynamic equations}
 
The radial, azimuthal and vertical components of the Navier-Stokes momentum equation are, respectively:
\beq{iur}
\mathrm{i}\omega u_r-2\Omega u_\phi=\mathrm{i}k_r\frac{p_1}{\rho_0}+\frac{\rho_1}{\rho_0^2}\frac{\partial
p_0}{\partial r}-\nu k_r^2u_r+\nu\frac{\partial^2 u_r}{\partial z^2}\,,
\eeq

\beq{iuphi}
\mathrm{i}\omega u_\phi+\frac{\varkappa^2}{2\Omega}u_r=-\nu k_r^2u_\phi+
\nu\frac{\partial^2 u_\phi}{\partial z^2}\,,
\eeq

\beq{iuz}
\mathrm{i}\omega u_z=-\frac{1}{\rho}\frac{\partial p_1}{\partial z}+\frac{\rho_1}{\rho_0^2}\frac{\partial p_0}{\partial z}-\nu k_r^2u_z +
\nu\frac{\partial^2 u_z}{\partial z^2}
\eeq

Here 
\beq{}
\varkappa^2=4\Omega^2+r\frac{d\Omega^2}{dr}\equiv \frac{1}{r^3}\frac{d\Omega^2r^4}{dr}
\eeq
is the epicyclic frequency. For the power-law rotation $\Omega^2\sim r^{-q}$ the 
epicyclic frequency is simply $\varkappa^2/\Omega^2=4-q$.

In deriving these equations we have set to unity
the correction factors  $[R],[\Phi],[Z], [E] $ introduced in Paper I and which
take into account the dependence of the viscosity coefficient 
on temperature $\eta\sim T^{\alpha_{visc}}$ ($\alpha_{visc}=5/2$ for fully ionized gas and $\alpha_{visc}=1/2$ for neutral gas) in the perturbed viscous force component ${\cal N}_r, {\cal N}_\phi, {\cal N}_z$, respectively  (see Appendix A and Eq. (42) in Paper I), because the deviations of these coefficients from unity 
have a very insignificant effect on the results.
Below we shall also neglect the second derivatives with respect to $z$ 
of the perturbed velocity components in the dynamical equations \eqn{iur}-\eqn{iuz}. 
This can be justified if $\nu k_r^2  u_{r,\phi,z}>\nu (\partial^2 u_{r,\phi,z}/\partial z^2)$. To within a numerical factor, this inequality 
can be recast to the form $(k_rr)^2 (z_0/r)^2> 1$. We shall see below that 
for $k_rr\gtrsim 100$ where the maximum instability increments occur
and for thin discs with $z_0/r\sim 0.02$ (see \Eq{z0r})
this is indeed the case.

\item{} \textbf{Pressure and entropy perturbations}

In the general case by varying the equation of state 
\Eq{e:eos} we obtain for entropy perturbations:
\beq{e:s1}
\frac{p_1}{p_0}=\frac{s_1}{c_V}+\gamma\frac{\rho_1}{\rho_0}\,.
\eeq
On the other hand, from the equation of state for ideal gas in the form 
$p=\rho {\cal R} T/\mu$, we find for small temperature perturbations  
we have:
\beq{e:T1}
\frac{p_1}{p_0}=\frac{\rho_1}{\rho_0}+\frac{T_1}{T_0}\,.
\eeq

\item{} \textbf{Energy equation}

The linearised viscous dissipation function is  
\beq{Qv}  
Q_\mathrm{visc}=\nu\rho r \frac{d \Omega}{d r}\left[r\frac{d\Omega}{d
r}-2\mathrm{i}k_r u_{\phi}-2\frac{u_{\phi}}{r}\right] + \hbox{quadratic\, terms}\,.
\eeq 
Here $\Omega=u_{\phi,0}/r$ is the angular (Keplerian) velocity of the unperturbed flow. The 
linearised energy equation takes the form 
\beq{e:en1}
 \frac{\rho_0{\cal R} T_0}{\mu}
\left(\mathrm{i}\omega s_1+
u_z\frac{\partial s_0 }{\partial z}+u_r\frac{\partial s_0 }{\partial r}
\right)
=-2\mathrm{i}k_r \nu\rho_0 r \frac{d \Omega}{d r}u_{\phi}
-\kappa k_r^2T_0\frac{T_1}{T_0}\,,
\eeq

\end{enumerate}

To take into account the heat conductivity effects, 
it is convenient to introduce the dimensionless Prandtl number:
\beq{e:Pre}
\mathrm{Pr}\equiv \frac{\nu\rho_0 C_p}{\kappa}=\frac{\nu\rho_0({\cal R}/\mu) c_p}{\kappa}=
\frac{\nu \rho_0 ({\cal R}/\mu)}{\kappa}\frac{\gamma}{\gamma-1}\,.
\eeq
The Prandtl number defined by \Eq{e:Pre} for fully ionized hydrogen 
gas ($\gamma=5/3$), where the heat conduction is determined by light electrons, is quite low (see \cite{1962pfig.book.....S}):
\beq{e:Pr_spitzer}
\mathrm{Pr}_\mathrm{e}\approx \frac{0.406}{20\cdot 0.4\cdot 0.225\cdot (2/\piup)^{\frac{3}{2}}}\myfrac{m_e}{m_p}^{1/2}\myfrac{5}{2}\approx 0.052\,.
\eeq

Note that the presence of magnetic field in plasma decreases both  
electron heat conductivity and viscosity. In this case 
both the viscosity and heat conductivity are determined by 
ions that have larger Larmor radius than electrons, and the Prandtl number
even in the case of fully ionized gas becomes 
\cite{1962pfig.book.....S}
\beq{}
\mathrm{Pr}_\mathrm{i}=\frac{3}{20}c_p
\eeq
which is 3/8 for $\gamma=5/3$.

In the case of cold neutral gas the Prandtl number is  
Pr$_n$=2/3 according to simplified kinetic theory 
\citep{Hirschfelder_al54}, and the heat conductivity 
coefficient depends on temperature as 
$\kappa\sim T^{1/2}$ \citep{1962pfig.book.....S}.

After eliminating the temperature variations in the energy equation using  \Eq{e:s1} and \Eq{e:T1}, we find
\beq{e:rho1uphi}
\frac{\rho_1}{\rho_0}\left(\mathrm{i}\omega  + \frac{\nu k_r^2}{\mathrm{Pr}}\right)
-\frac{1}{c_p}\left(u_z\frac{\partial s_0 }{\partial z}+u_r\frac{\partial s_0 }{\partial r}\right)=
\frac{2\mathrm{i}k_r \nu r (d \Omega/d r)}{c_p{\cal R}T_0/\mu}u_{\phi} +
\dfrac{p_1}{p_0}\left(\mathrm{i}\dfrac{\omega}{\gamma}+\dfrac{\nu  k_r^2}{\mathrm{Pr}}
\right)
\eeq 
Here $c_p=\gamma c_V=\gamma/(\gamma-1)$ is the specific heat capacity (per particle) at constant pressure. 

We will neglect very slow variations of the unperturbed pressure, density and
entropy along the radial coordinate, i.e. set $\partial /\partial r=0$ in 
the continuity equations \eqn{e:anel}, dynamic equations \eqn{iur}-\eqn{iuz} and 
energy equation \eqn{e:rho1uphi}. Below we shall also denote the partial
derivative with respect to $z$ by prime.
Thus we are left with the following system of five linearised hydrodynamic equations in the anelastic approximation for five 
variable $u_r, u_z, u_\phi, \rho_1/\rho_0, p_1/p_0$:
\beq{lcont}
u_z'-\mathrm{i}k_ru_r+\frac{\rho_0'}{\rho_0}u_z=0\,,
\eeq
\beq{lur}
(\mathrm{i}\omega +\nu k_r^2)u_r-2\Omega u_\phi=\mathrm{i}k_r\frac{p_1}{\rho_0}\,,
\eeq
\beq{luphi}
(\mathrm{i}\omega+\nu k_r^2)u_\phi+\frac{\varkappa^2}{2\Omega}u_r=0\,,
\eeq
\beq{luz}
(\mathrm{i}\omega +\nu k_r^2)u_z=-\frac{p_1'}{\rho_0}+\frac{\rho_1}{\rho_0}
\frac{p_0'}{\rho_0}\,,
\eeq
\beq{le}
\frac{\rho_1}{\rho_0}\left(\mathrm{i}\omega  + \frac{\nu k_r^2}{\mathrm{Pr}}\right)=
\frac{2\mathrm{i}k_r \nu r (d \Omega/d r)}{c_p{\cal R}T_0/\mu}u_{\phi}+\frac{1}{c_p}s_0'u_z \,.
\eeq
In the right-hand side of 
energy equation \eqn{le} we have omitted the term due to pressure perturbations $\propto p_1/p_0$ 
because otherwise it will give rise to spurious unstable modes in the case of 
steady solid-body rotation with $\Omega=const$.

\section{Derivation of the dispersion equation}
\label{s:dispeq}

We start with substituting $\rho_1/\rho_0$ from \Eq{le} into \Eq{luz}. Here in the right-hand side of the resulting equation two coefficients 
depending on the $z$-coordinate arise: 
\beq{Phi0}
\Phi_0\equiv \frac{p_0'}{p_0}\nu
\eeq
and the Brunt-V\"ais\"al\"a frequency:
\beq{BV}
-N_z^2\equiv\frac{p_0'}{\rho_0}\frac{s_0'}{c_p}=\frac{p_0'}{\rho_0}\frac{\partial}{\partial z}\myfrac{p_0^{1/\gamma}}{\rho_0}\,.
\eeq
After differentiating the resulting equation for $u_z$ with respect to $z$ and
eliminating $u_z'$, $u_r$ and $u_\phi$ using \eqn{lcont}-\eqn{luz}, we arrive 
at the following second-order linear differential equation for density perturbations
\beq{p''}
p_1''+Ap_1'+Bp_1=0
\eeq
with coefficients:
\beq{A}
A=\dfrac{(-N_z^2)'}{(\mathrm{i}\omega+\frac{\nu k_r^2}{\mathrm{Pr}})(\mathrm{i}\omega+\nu k_r^2)-
(-N_z^2)}-\Phi_0\dfrac{\varkappa^2}{(\mathrm{i}\omega+\nu k_r^2)^2}
\dfrac{(d\ln\Omega/d\ln r)}{c_p(\mathrm{i}\omega+\frac{\nu k_r^2}{\mathrm{Pr}})}
\dfrac{k_r^2}{1+\dfrac{\varkappa^2}{(\mathrm{i}\omega+\nu k_r^2)^2}}
\eeq
\beq{B}
B=\dfrac{-k_r^2}{1+\dfrac{\varkappa^2}{(\mathrm{i}\omega+\nu k_r^2)^2}}
\left[1-\dfrac{(-N_z^2)}
{(\mathrm{i}\omega+\nu k_r^2)(\mathrm{i}\omega+\frac{\nu k_r^2}{\mathrm{Pr}})}
+\Phi_0'\dfrac{\varkappa^2}{(\mathrm{i}\omega+\nu k_r^2)^2}
\dfrac{(d\ln\Omega/d\ln r)}{c_p(\mathrm{i}\omega+\frac{\nu k_r^2}{\mathrm{Pr}})}
\left(
1+\mydfrac{\Phi_0}{\Phi_0'}
\dfrac{(-N_z^2)'}{(\mathrm{i}\omega+\frac{\nu k_r^2}{\mathrm{Pr}})(\mathrm{i}\omega+\nu k_r^2)-
(-N_z^2)}
\right)
\right]
\eeq

Let us introduce a new variable
\beq{}
p_1=Ye^{-\frac{1}{2}Az}
\eeq
to eliminate the first derivative term in \Eq{p''}:
\beq{SL}
Y''+(B-\dfrac{1}{4}A^2)Y=0\,.
\eeq
This equation should be supplemented with two boundary conditions: 
\beq{bc1}
Y_1(z_0)=0, \quad Y_1(0)=0
\eeq
or 
\beq{bc2}
Y_2(z_0)=0, \quad Y_2'(0)=0
\eeq
for two linearly independent solutions. Using the form of the 
coefficient $A$ \eqn{A} it is easy to check that these boundary conditions
exactly correspond to the physically motivated boundary conditions for pressure variations: $p_1(z_0)=0$, $p_1(0)=0$ and $p_1(z_0)=0$, $p_1'(0)=0$, respectively.

This Sturm-Liouville problem \eqn{SL}-\eqn{bc2} can be easily
solved if coefficients $A$ and $B$ are independent of $z$. Therefore, it is necessary to  average these coefficients over some assumed background disc vertical structure. As the background solution, we will use the polytrope 
discs with (see \cite{1998A&AT...15..193K}):
\beq{}
\rho_0(z)=\rho_c\left(1-\myfrac{z}{z_0}^2\right)^n, \quad p_0(z)=p_c\left(1-\myfrac{z}{z_0}^2\right)^{(n+1)}, \quad T_0(z)=T_c\left(1-\myfrac{z}{z_0}^2\right)\,.
\eeq
Here $n$ is the polytrope index, $\rho_c, p_c$ and $T_c$ are the values of 
density, pressure and temperature in the disc symmetry plane, respectively.
Therefore, the density-averaged values of quantities $\Phi_0, \Phi_0'$ and
$(-N_z^2)$ should be determined as 
\beq{}
\langle(...)\rangle\equiv \frac{\int_0^{z_0}(...)\rho_0(z)dz}{\int_0^{z_0}\rho_0(z)dz}
\eeq
where $z_0$ is the semi-thickness of the disc. Thus we obtain:
\beq{}
\Phi_0=\langle \Phi_0\rangle \frac{\Omega a r^2}{z_0}\,,
\eeq
where, as in Paper I, we have introduced the dimensionless viscosity parameter
$a$ through the free-path length of particles $l/r$ and the ratio of the sound velocity to the unperturbed angular (Keplerian) velocity $u_s/u_\phi$ 
\beq{a}
\frac{\nu k_r^2}{\Omega}=a(k_rr)^2\,, \quad
a\equiv\myfrac{u_s}{u_{\phi}} \myfrac{l}{r}\,.
\eeq
Note that the maximum possible mean-free path in thin discs 
in the frame of hydrodynamic treatment should be less than the disc thickness 
$z_0$, i.e. $l/r=(l/z_0)(z_0/r)\simeq (l/z_0) (u_s/u_\phi)<(u_s/u_\phi)$.
The derivative of $\Phi_0$ can be written as 
\beq{}
\Phi_0'=\langle \Phi_0'\rangle \frac{\Omega a r^2}{z_0^2}\,,
\eeq
The corresponding dimensionless mean values are 
\beq{}
\langle \Phi_0\rangle=-\dfrac{n+1}{\alpha_{visc}\langle\Sigma_0\rangle}
\eeq
\beq{}
\langle \Phi_0'\rangle=\dfrac{1}{\langle \Sigma_0\rangle}\left(
2(n+1)(n+2)B(\frac{3}{2},\alpha_{visc}-1)+\dfrac{n(n+1)}{\alpha_{visc}-1}
-2(n+1)nB(\frac{3}{2},n-1)
\right)\,.
\eeq
 Similarly, for the mean Brunt-V\"ais\"al\"a frequency we find:
\beq{}
\langle -N_z^2 \rangle=\dfrac{2\Omega^2}{\langle \Sigma_0\rangle}\left(\dfrac{n+1}{\gamma}-n\right)B(\frac{3}{2},n)\,.
\eeq
In the above formulas the dimensionless surface density of the disc is
\beq{}
\langle \Sigma_0\rangle\equiv \dfrac{\int_0^{z_0}\rho_0dz}{\rho_c z_0}=2^{2n}B(n+1,n+1)\,,
\eeq
$B(x,y)$ is the beta-function. 
When deriving these values, we have used the property that the microscopic 
dynamic viscosity 
is a function of temperature only $\nu_0\rho_0\sim T^{\alpha_{visc}}$, with 
$\alpha_{visc}=5/2$ for fully ionized gas. For neutral gas $\alpha_{visc}=1/2$ 
and the averaging should be performed with weight $\rho_0^2$ (see below
in Section \ref{s:neutralgas}). 


The solution of the Sturm-Liouville problem \Eq{SL} with boundary conditions
\eqn{bc1} and \eqn{bc2} results in eigenfunctions $\sin(\lambda_sz)$ and $\cos (\lambda_c z)$ and eigenvalues $\lambda_s=m\piup/z_0$ and $\lambda_c=(m-1/2)\piup/z_0$ ($m=1,2,...$)), respectively. Substituting the eigenfunctions into \Eq{SL} yields
the sought for dispersion equation:
\beq{}
-\lambda_{s,c}^2+B-\frac{1}{4}A^2=0\,.
\eeq

The form of the coefficient $A$ \eqn{A} immediately implies that 
the dispersion equation will be a tenth-order algebraic equation for $\omega$
with complex coefficients. It can be checked that 
for any value of $n>3/2$ (i.e. for 
convectively stable disc structure), two spurious 
unstable modes arise in the case of steady-state solid-body rotation (with $q=0$),
which disappear if we omit 
the terms with $(-N_z^2)'$ (note that this term is not dangerous 
for adiabatic background structure where $(-N_z^2)$ and its derivative 
vanish). Similarly unstable modes would arise if we retain 
pressure perturbations in the right-hand side of the 
energy equation \eqn{e:rho1uphi}. Due to subtleties with the 
energy equation in the Boussinesq and anelastic approximations mentioned above 
and analysed in \citep{2013ApJ...773..169V}, we conclude that 
the Brunt-V\"ais\"al\"a frequency should be kept constant
when differentiating \Eq{luz}. 
This might indicate that the background entropy gradient $s_0'$
should be omitted in the right-hand side of the 
energy equation \eqn{le} from  the very beginning. Not so, 
since its inclusion leads to the physically correct result (stabilization of
the modes) already in the Boussinesq approximation (see Paper I). 
Another possibility might be the keeping both $\Phi_0$ and $(-N_z^2)$ 
constant when deriving equation \eqn{p''} for pressure perturbations. Not so again,
since the factor $\Phi_0$ \eqn{Phi0} stems from the fully leguitime 
linearised form of $z-$component of the Navier-Stokes equation \eqn{iuz}.
While, of course, rigorous proof of such a treatment is highly 
desirable, here we restrict ourselves to the qualitative arguments given above.
Therefore, after crossing out terms with $(-N_z^2)'$ in \Eq{A} and \Eq{B}, 
we will be left with a sixth-order algebraic equation for $\omega$.

The found eigenfunctions for the variable $Y$ means that the 
eigenfunctions for the pressure and velocity perturbations 
have the form $\exp(-Az/2)\cos(\lambda_s z)$ or $\exp(-Az/2)\sin(\lambda_s z)$, 
which is typical for perturbations in stratified
atmospheres \cite{2013ApJ...773..169V}.
We will find that the maximum increment is reached for the $\cos$ mode, so
substituting coefficients $A$ and $B$ from \Eq{A} and \Eq{B} yields the 
following quantized dispersion equation:
\begin{eqnarray}
\label{dispeq}
&-\mydfrac{(m-1/2)\piup}{z_0}^2-
\dfrac{k_r^2}{\left(1+\dfrac{\varkappa^2}{(\mathrm{i}\omega+\nu k_r^2)^2}\right)}
\left[1-\dfrac{(-N_z^2)}
{(\mathrm{i}\omega+\nu k_r^2)(\mathrm{i}\omega+\frac{\nu k_r^2}{\mathrm{Pr}})}
+\Phi_0'\dfrac{\varkappa^2}{(\mathrm{i}\omega+\nu k_r^2)^2}
\dfrac{(d\ln\Omega/d\ln r)}{c_p(\mathrm{i}\omega+\frac{\nu k_r^2}{\mathrm{Pr}})}
\right]\nonumber \\
&-\dfrac{1}{4}\left[
-\Phi_0\dfrac{\varkappa^2}{(\mathrm{i}\omega+\nu k_r^2)^2}
\dfrac{(d\ln\Omega/d\ln r)}{c_p(\mathrm{i}\omega+\frac{\nu k_r^2}{\mathrm{Pr}})}
\dfrac{k_r^2}{\left(1+\dfrac{\varkappa^2}{(\mathrm{i}\omega+\nu k_r^2)^2}\right)}
\right]^2=0\,.
\end{eqnarray}

This is a sixth-order algebraic equation with complex coefficients (cf. cubic dispersion
equation (31) from Paper I derived in the Boussinesq limit using local WKB-analysis).

\section{Solution of the dispersion equation}
\label{s:solution}

Let us analyse the solution of the dispersion equation in the anelastic 
approximation derived above for two fluids: the case of fully ionized gas
with the Prandtl number Pr$_{e}=0.053$ and Pr$_{i}=3/8$ if the magnetic field is 
present, and the case of neutral gas with Pr$_n=2/3$. The last case should be
treated separately, since for the dependence of the dynamical viscosity $\propto 
T^{1/2}$ the averaging over the vertical coordinate 
with the weight $\rho_0(z)$ is insufficient -- near the surface
layers of the disc the mean free-path length of particles 
is so large that leads to divergences in the term $\Phi_0'$.

We will consider the laminar shear flow with the velocity 
profile $\Omega^2\propto r^{-q}$ so that the shear coefficient is 
$d\ln\Omega/d\ln r=-q/2$. All relevant frequencies will be normalized to the 
local Keplerian value $\Omega$ and denoted with tilde. In the numerics 
the adiabatic index of the gas is set to $\gamma=5/3$.

It is convenient to write down 
the dimensionless dispersion equation for the dimensionless 
mode frequency $\tilde\omega$ as a function of the dimensionless variable 
$(k_rr)$ with dimensionless viscosity parameter $a$ and 
the disc thickness 
\beq{z0r}
\frac{z_0}{r}=\sqrt{\frac{\Pi_1}{\gamma}}\myfrac{u_{s}}{u_\phi}\,.
\eeq
Here the dimensionless factor $\Pi_1$ takes into account the vertical disc structure, and in the case of the polytrope accretion 
discs $\Pi_1=2(n+1)$ \citep{1998A&AT...15..193K}.
The values of the sound velocity $u_s$ then should be taken in the disc symmetry plane.
 
The dispersion equation \eqn{dispeq} in the dimensionless form reads:
 
\begin{eqnarray}
\label{dispeq1}
&\mydfrac{(m-1/2)\piup}{z_0/r}^2+
\dfrac{(k_rr)^2}{\left(1+\dfrac{\tilde \varkappa^2}{(\mathrm{i}\tilde\omega+ a(k_rr)^2)^2}\right)}
\left[1-\dfrac{\langle -\tilde N_z^2\rangle}
{(\mathrm{i}\tilde\omega+a (k_rr)^2)(\mathrm{i}\tilde\omega+\frac{a(k_rr)^2}{\mathrm{Pr}})}
+\dfrac{\langle\Phi_0'\rangle a}{(z_0/r)^2}\dfrac{\tilde\varkappa^2}{(\mathrm{i}\tilde\omega+a (k_rr)^2)^2}
\dfrac{(-q/2)}{c_p(\mathrm{i}\tilde\omega+\frac{a (k_rr)^2}{\mathrm{Pr}})}
\right]\nonumber \\
&+\dfrac{1}{4}\left[
\dfrac{\langle\Phi_0\rangle a}{(z_0/r)}\dfrac{\tilde\varkappa^2}{(\mathrm{i}\tilde\omega+a (k_rr)^2)^2}
\dfrac{(-q/2)a}{c_p(\mathrm{i}\tilde\omega+\frac{a (k_rr)^2}{\mathrm{Pr}})}
\dfrac{(k_rr)^2}{\left(1+\dfrac{\tilde\varkappa^2}{(\mathrm{i}\tilde\omega+a(k_rr)^2)^2}\right)}
\right]^2=0\,.
\end{eqnarray}

\subsection{Fully ionized gas}

\begin{figure*}
\begin{center}
\includegraphics[width=0.48\textwidth]{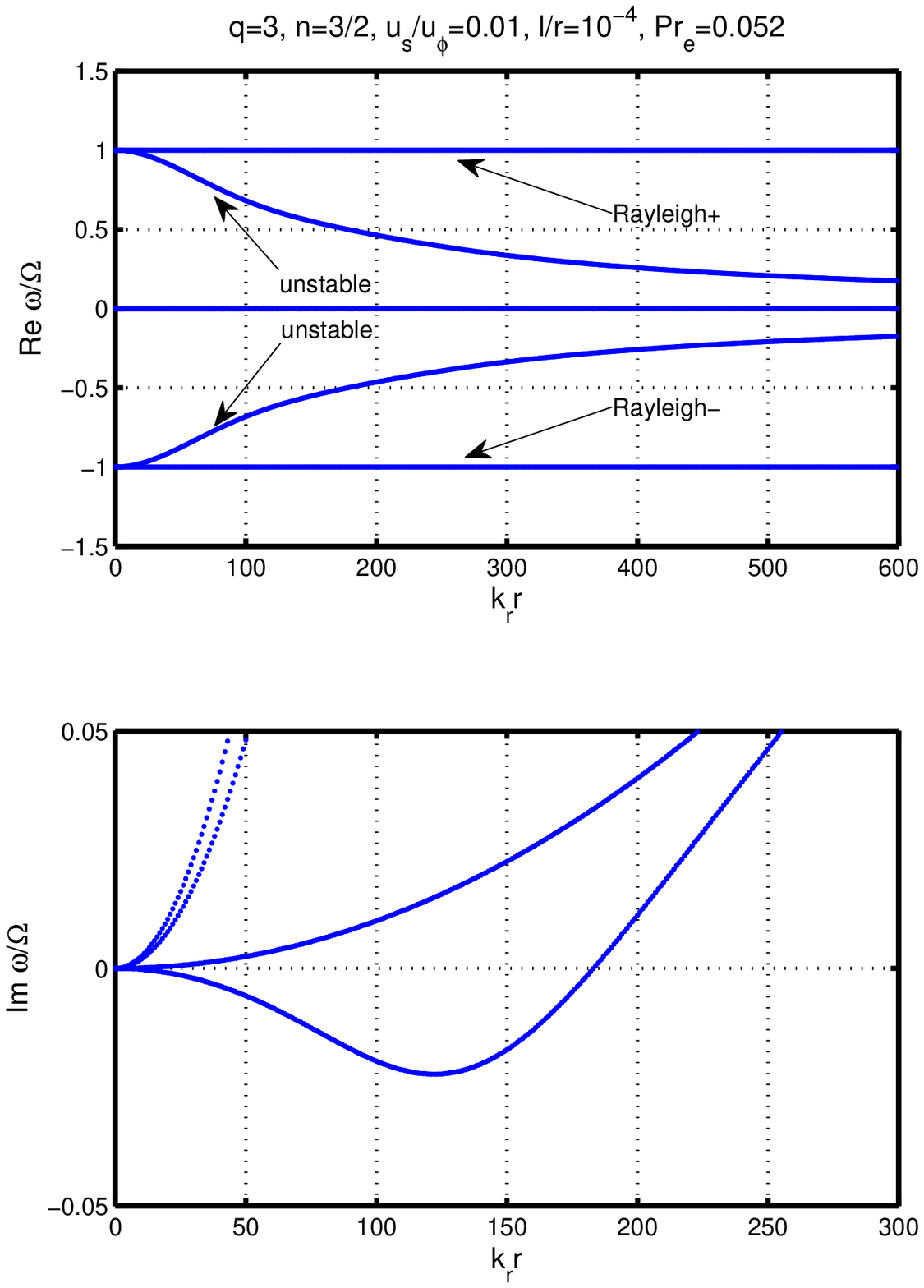}
\hfill
\includegraphics[width=0.48\textwidth]{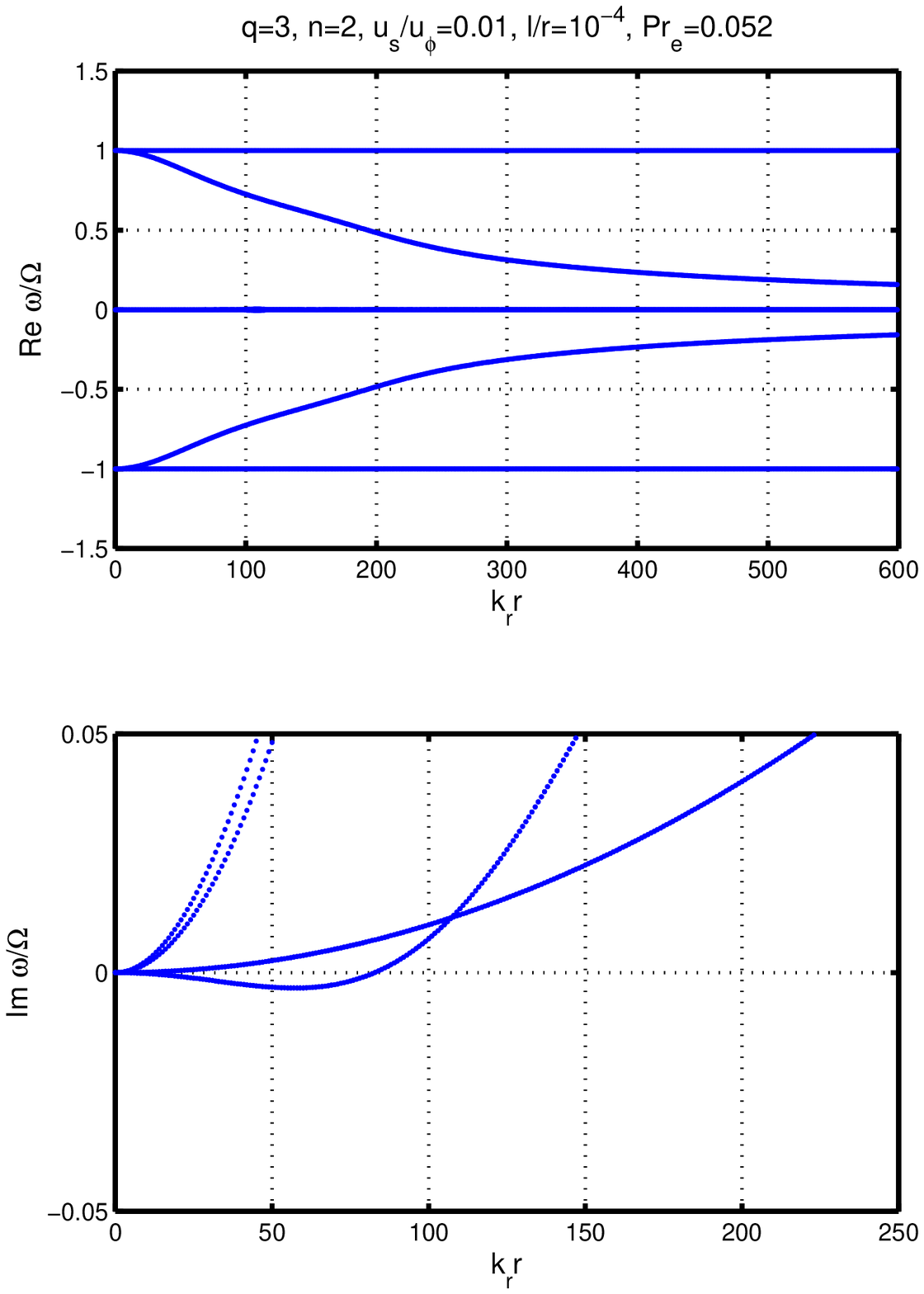}
\caption{Left: Real and imaginary parts of anelastic modes in fully ionized gas
with electron heat conductivity (Pr$_e$=0.052) and viscosity parameters 
$u_s/u_\phi=0.01, l/r=10^{-4}$ for the adiabatic density distribution ($n=\frac{3}{2}$).
Right: The same for vertically increasing entropy distribution with 
$n=2$.}
\label{f:ionized}
\end{center}
\end{figure*}

The results of the solution of the dispersion equation \eqn{dispeq1} are shown 
in Fig. \ref{f:ionized} for two background vertical structures of a thin Keplerian disc. In the left panel of Fig. \ref{f:ionized} we present the solution for the polytrope disc structure with constant entropy described by the polytrope index $n=\frac{3}{2}$. Here the Brunt-V\"as\"al\"a frequency $N_z^2$ vanishes. It is seen that the unstable 
anelastic mode (the one with the negative imaginary part in the bottom panel) arises at $k_r\sim 50-150$, where the approximation of the incompressibility
($\partial \rho/\partial t=0$) is justified.
In fact, this is two modes with equal absolute values 
but different by sign real parts that 
demonstrate the overstability (marked with arrows in the upper left panel). This 
is different from the Boussinesq limit considered in Paper I, where 
one of the Rayleigh inertial modes became viscously overstable in the presence of viscosity. In the anelastic approximation, the Rayleigh inertial modes remain always stable, and the unstable modes are split out from the Rayleigh modes. In the right panels of Fig. \ref{f:ionized} we present the solutions for the background disc structure with vertically 
increasing entropy (shown is the solution for $n=2$), which is convectively stable ($N_z^2>0$) in the absence of viscosity and shear. However, the presence of even small viscosity makes the Keplerian flow convectively unstable even in this case.      

\begin{figure}
\begin{center}
\includegraphics[width=0.33\textwidth]{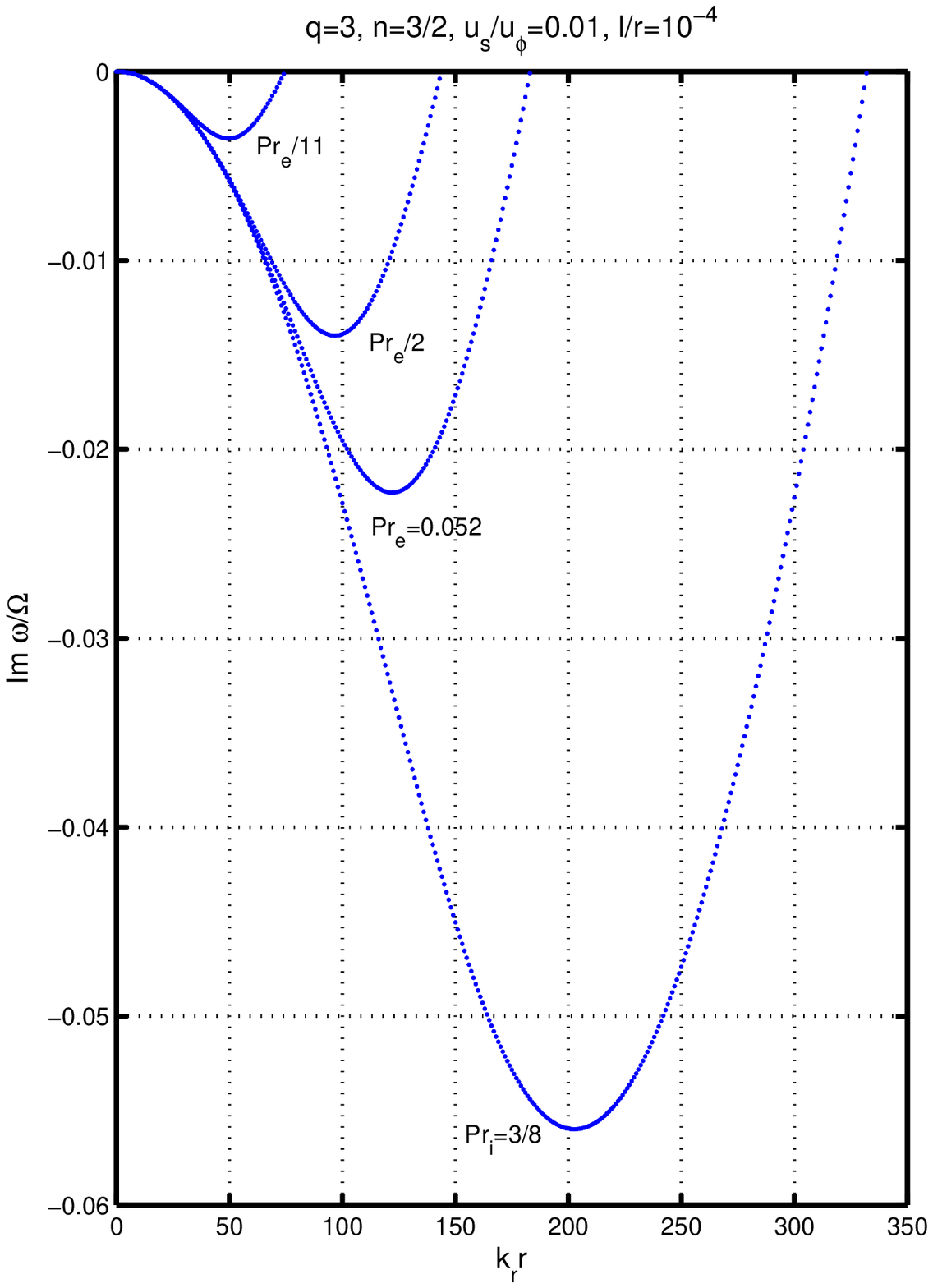}
\hfill
\includegraphics[width=0.33\textwidth]{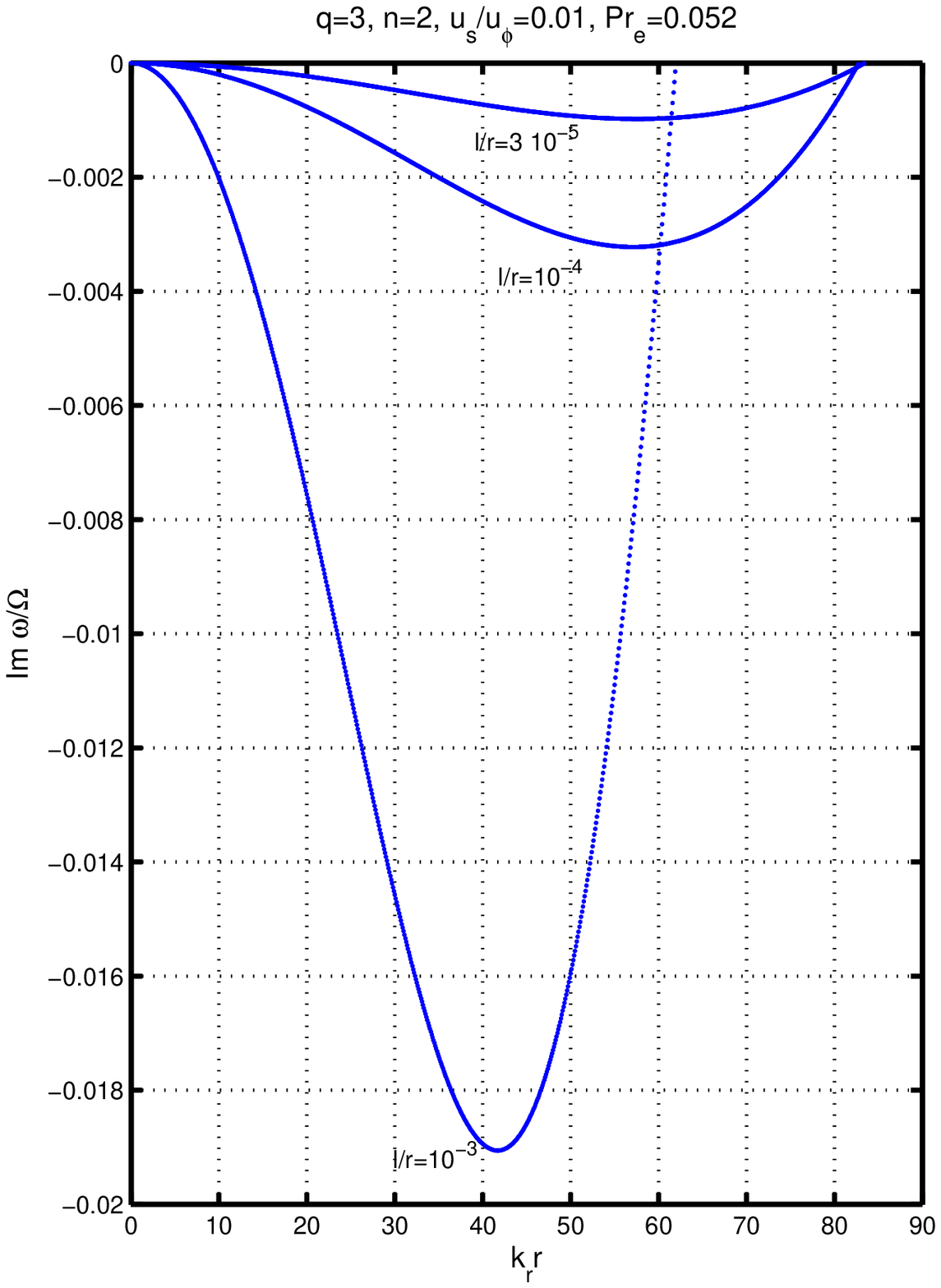}
\hfill
\includegraphics[width=0.33\textwidth]{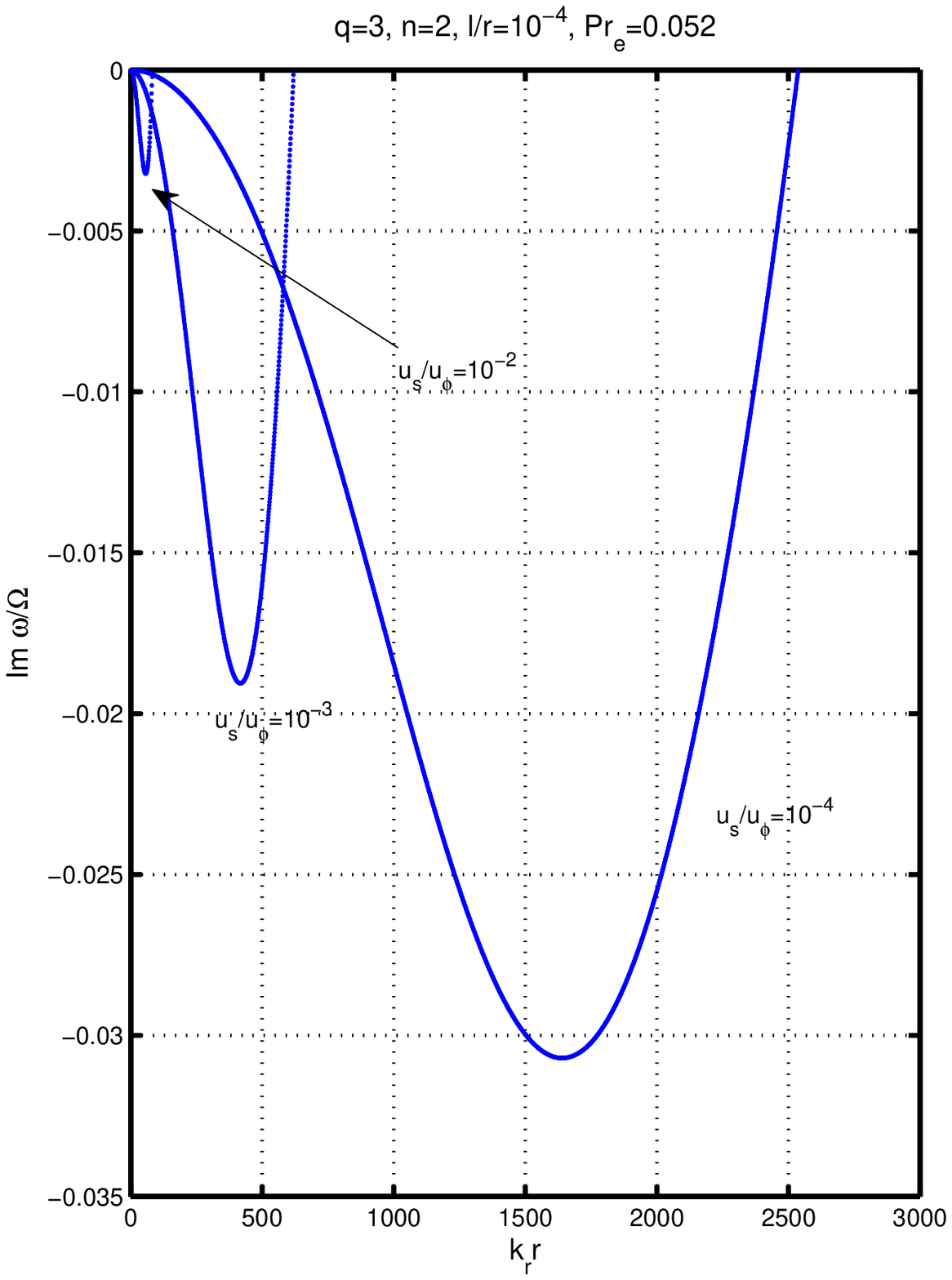}
\caption{Left: Imaginary part of the viscously unstable anelastic mode in fully ionized gas
with different Prandtl numbers. Viscosity parameters 
$u_s/u_\phi=0.01, l/r=10^{-4}$ for vertically increasing entropy distribution with $n=2$. Middle: The same for different mean free-path length of particles $l/r$ with fixed
disc thickness parameter $u_s/u_\phi=0.01$. Right: The same with different disc thickness parameter $u_s/u_\phi$ for fixed mean free-path length of particles $l/r=10^{-4}$.}
\label{f:imag}
\end{center}
\end{figure}

In Fig. \ref{f:imag} we explore the effect of different dimensionless parameters of 
the problem on the increment of the overstability. First, in the left panel of Fig. \ref{f:imag} we study the effect of changing  
the Prandtl number, which describes
the dumping effect of thermal conductivity on small perturbations. For fully ionized gas, the Prandtl number is maximal when the small (but still dynamically unimportant) magnetic field is present, and both the viscosity and heat conductivity 
are mediated by ions which have larger Larmor radius than electrons (Pr$_{i}$=3/8, see above). We also show the results of the dumping effect of possible radiative conductivity parameterized in terms of the effective Prandtl number (Pr/2 and Pr/11, see
Eq. (47) in Paper I). The smaller the Prandtl number, the smaller the instability increment, which is physically clear.  Second, in the central panel of Fig.  \ref{f:imag} we show the effect of changing the viscosity (parametrized in terms of 
the effective mean free-path length of particles, $l/r$). The larger the viscosity, the
higher the instability increment. Finally, in the left panel of Fig. \ref{f:imag}
we demonstrate the effect of the disc semi-thickness (parametrized by the central 
sound speed to the angular velocity ratio, $u_s/u_\phi$). At a given mean-free path of particles, the thinner the disc, the higher the increment.

\subsection{Neutral gas}
\label{s:neutralgas}

For fully neutral gas with the Prandtl number Pr$_n=2/3$ we should first 
make new averaging of quantities $\Phi_0$, $\Phi_o'$ and $(-N_z^2)$ with 
weight $\rho_0^2(z)$ to avoid divergence of the value $\Phi_0'$ near the disc surface due to very large mean-free path length of particles in the polytropic discs. 
We find:
\beq{}
\langle\langle \Phi_0\rangle\rangle=-\dfrac{n+1}{(\alpha_{visc}+n)\langle\langle \Sigma_0\rangle\rangle}
\eeq
\beq{}
\langle\langle \Phi_0'\rangle\rangle=
\dfrac{1}{\langle\langle \Sigma_0\rangle\rangle}\left(
2(n+1)(n+2)B(\frac{3}{2},\alpha_{visc}-1+n/2)+\dfrac{n(n+1)}{\alpha_{visc}-1+n}
-2(n+1)nB(\frac{3}{2},n-1+n/2)\right)
\eeq
\beq{}
\langle\langle (-N_z^2)\rangle \rangle=\dfrac{2}{\langle\langle \Sigma_0\rangle\rangle}\left(\dfrac{n+1}{\gamma}-n\right)B(\frac{3}{2},2n)\,.
\eeq
\beq{}
\langle\langle \Sigma_0\rangle\rangle
=2^{4n}B(2n+1,2n+1)\,,
\eeq

\begin{figure*}
\begin{center}
\includegraphics[width=0.48\textwidth]{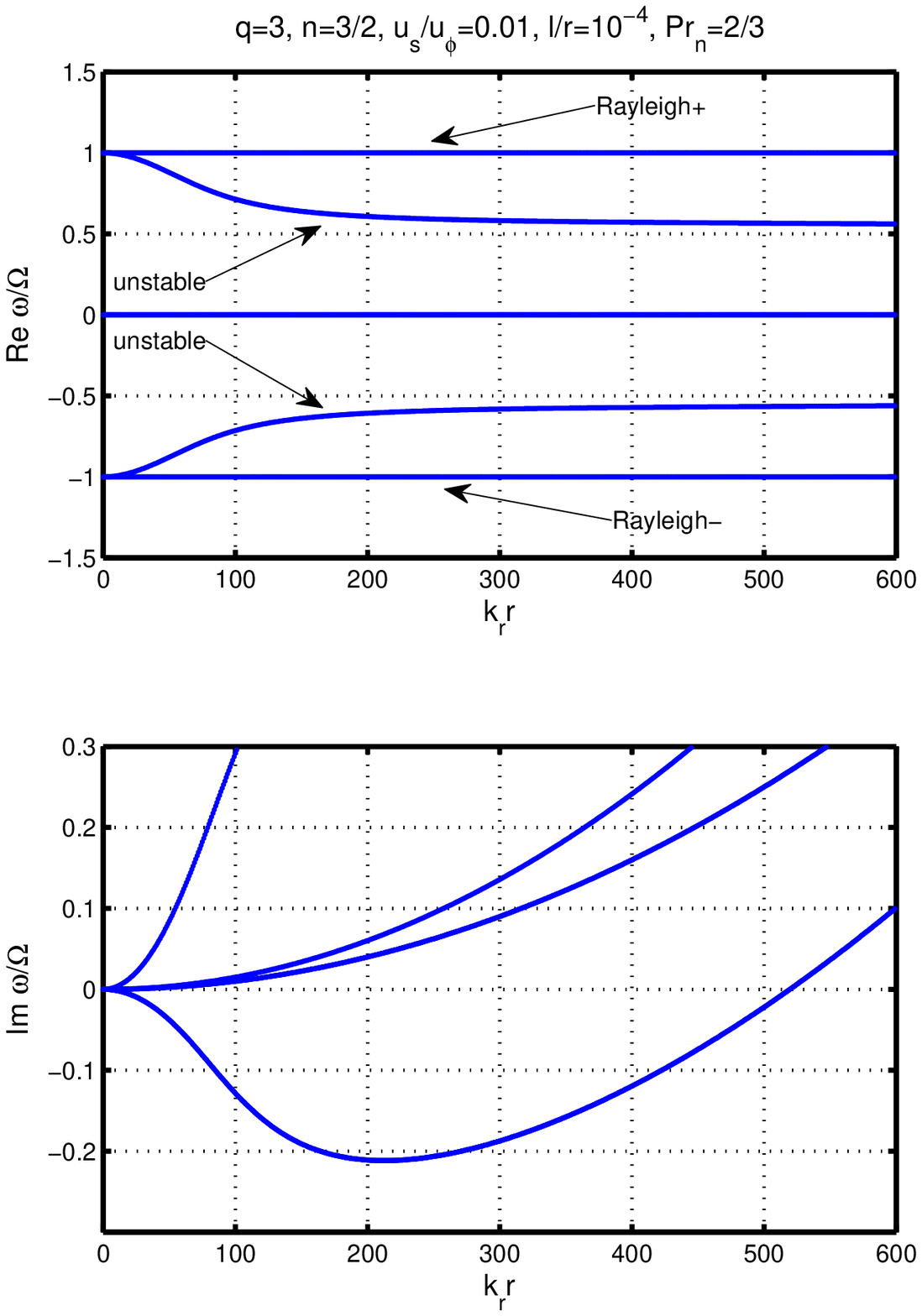}
\hfill
\includegraphics[width=0.48\textwidth]{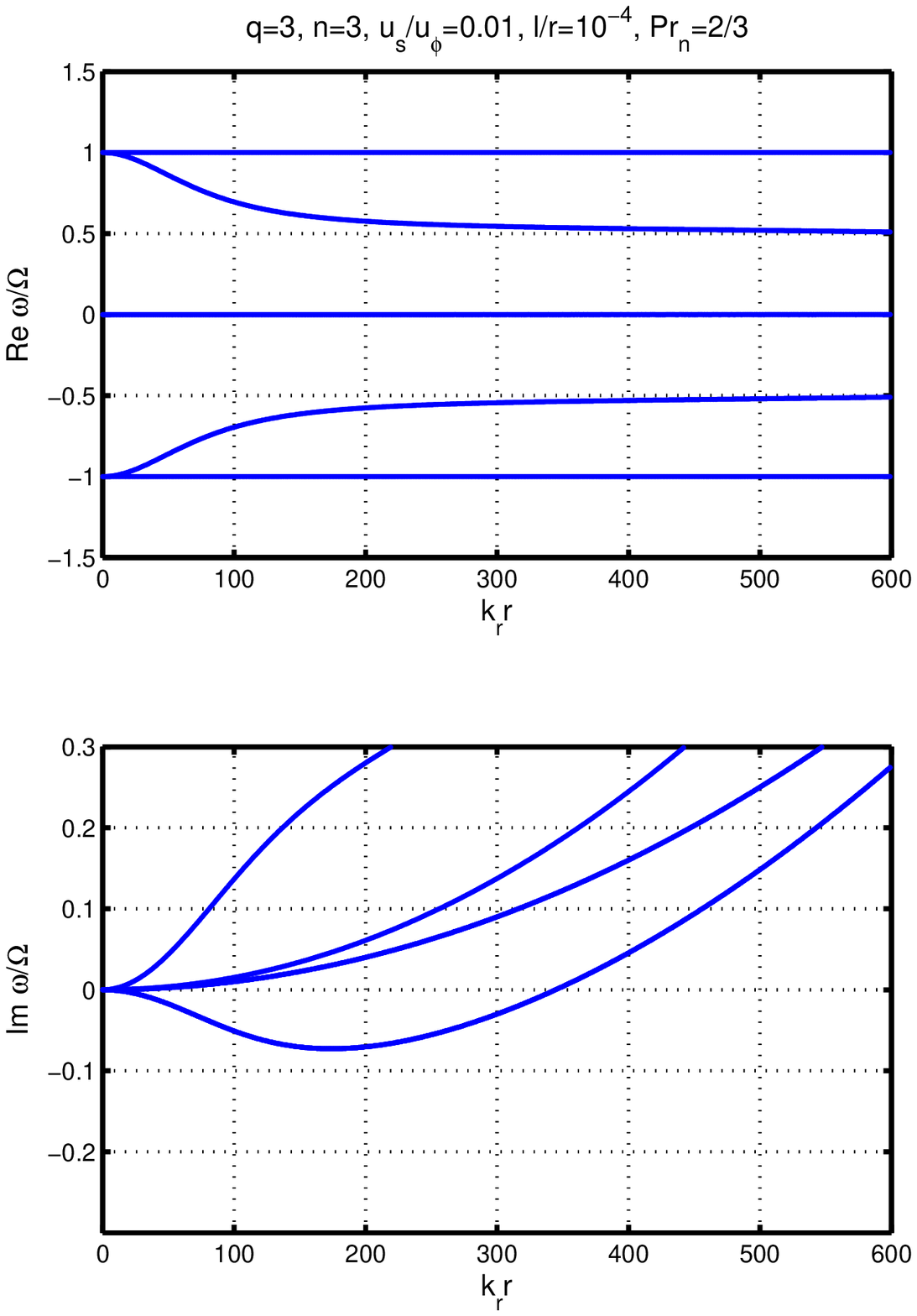}
\caption{Left: Real and imaginary parts of anelastic modes in neutral gas
for the adiabatic entropy distribution ($n=\frac{3}{2}$).
Right: The same 
for vertically increasing entropy distribution with $n=3$. Note that the real part of the solution is quite insensitive to the polytropic structure of the disc.}
\label{f:neutral}
\end{center}
\end{figure*}

In Fig. \ref{f:neutral} we show the real and imaginary parts of anelastic modes 
for the case of neutral gas with the Prandtl number Pr$_n=2/3$.
The disc thickness is $z_0/r\sim u_s/u_\phi=0.01$, the mean free path of ions is $l/r=10^{-4}$.
 Since the Prandtl number is quite large, unlike the case of 
the ionized gas shown in Fig. \ref{f:ionized}, 
the instability is present even for the quite significant 
vertical background entropy gradient ($n=3$, the right panel of this Figure). 

\section{Discussion and conclusions}
\label{s:discussion}

In the present paper we have extended the modal analysis of small 
axially symmetric perturbation in sheared Keplerian flows with microphysical viscosity and heat conductivity, which we started in Paper I.
In contrast to Paper I, where the Boussinesq approximation was used, here we 
have formulated the problem in the anelastic approximation and taken into account
vertical boundary conditions in thin discs. In this approximation we have obtained the second-order linear differential equation with respect to the $z$-coordinate for small pressure perturbations, \Eq{p''}, with coefficients (\Eq{A} and \Eq{B}) depending on 
the height above the disc plane $z$. We have assumed the background polytropic vertical disc structure and averaged the coefficients  \Eq{A} and \Eq{B} with weight $\rho_0(z)$ over the vertical disc height. This allowed us to solve the Sturm-Liouville problem to obtain the discrete spectrum of eigenvalues and eigenfunctions ($\cos(\lambda_c z)$ or $\sin(\lambda_s z)$). Substitution of 
these functions into \Eq{p''} resulted in the dispersion equation for normal modes 
of axially symmetric perturbations along the radial coordinate, \Eq{dispeq}.
This turned out to be an algebraic sixth-order equation, solution of which for a wide 
range of the viscosity parameters in thin Keplerian discs are presented in Fig. \ref{f:ionized}-\ref{f:neutral}. 
Note that in the Boussinesq limit the dispersion equation had only the third order.
We have found that in a wide range of 
wavenumbers $k_rr\gg1 $ two unstable modes are split from the classical 
inertial Rayleigh modes (in the Boussinesq limit one of the Rayleigh modes
displayed the overstability in the presence of viscosity). Qualitatively, the results of the present 
paper are in agreement with findings
of Paper I. However, in contrast to Paper I, where the Boussinesq approximation was used, the results are quantitatively different.

The description of the hydrodynamic flows in the anelastic approximation, although neglects the sound modes, 
is more precise and takes into account the important term in the continuity equation, $(1/\rho_0\partial \rho_0/\partial z)u_z$. This means that the analysed small perturbations are no more purely transversal, as in the Boussinesq limit. 
However, it is important to note that in both anelastic and Boussinesq approximations the pressure variations $p_1/p_0$ 
should be neglected in the energy equation, otherwise fictitious unstable solutions 
emerge in the case of the steady-state solid-body rotation. 
We have also found that the overstability appears in the cases where 
there is a non-zero 
vertical gradient of the quantity $(p_0'/p_0)\nu\sim -(\Omega^2 z/T_0)\nu$.

In both approximation we
have found that the increment of overstable modes increases with viscosity and 
the background vertical pressure gradient, suggesting a convective nature of the 
overstability: the viscous heat generation in the sheared flow in the gravitational field around a central star makes the flow convectively unstable. It is tempting to suggest that this instability may 
be the seed for turbulence in Keplerian discs even in the absence of magnetic fields.    
 
Note the many faces of the viscosity in the considered problem.
The higher the viscosity in the right-hand side of energy equation \eqn{le}, 
the stronger the viscous energy generation due to the shear 
leading to the buoyancy of the 
perturbed regions. On the other hand, in the dynamic equations \eqn{iur}-\eqn{iuz}
the viscosity damps the perturbations. In the right-hand side of 
these equations, we have neglected 
terms $\nu u_r'', \nu u_\phi'', \nu u_z''$ with second derivatives of 
perturbed velocities. We expect that their taking into account 
will somewhat decrease the increment of the viscous-convective instability and 
narrow the interval of unstable wavenumbers in Figs. \ref{f:ionized} and \ref{f:neutral}. If we retain these second derivatives, we will obtain 
a much more complicated sixth-order differential equation for perturbations.
This is a separate problem to be addressed elsewhere. Here we have restricted
ourselves to solving only the second-order differential equation \Eq{p''}.

\section*{Acknowledgements}

We thank Drs. Vladimir N. Lukash, Pavel B. Ivanov and other participants of the ASC FIAN theory department seminar for useful discussions.    
We are grateful to the anonymous referee for 
drawing our attention to the recent analysis of anelastic approximation 
by \cite{2013ApJ...773..169V}.
The work is supported by the Russian Science Foundation grant 14-12-00146.

\bibliographystyle{mn2e}
\expandafter\ifx\csname natexlab\endcsname\relax\def\natexlab#1{#1}\fi
\bibliography{mri}


\end{document}